%% file: main.tex
\documentclass[sigconf]{acmart}

\usepackage{graphicx}
\usepackage{booktabs}
\usepackage{multirow}
\usepackage{tabularx}
\usepackage{enumitem}
\usepackage{amsmath}
\usepackage{url}

\AtBeginDocument{%
  
}

\renewcommand\footnotetextcopyrightpermission[1]{}
\begin{document}


\title[Toward a Culturally Adapted Chinese Language Agent]{Toward a Culturally Adapted Chinese Language Agent: A Wizard-of-Oz Study of Nonverbal Behavior in Chinese-German Intercultural Interaction}


\author{Siddhant Jain}
\orcid{0009-0005-3051-7064}
\affiliation{%
  \institution{Deutsches Forschungszentrum f{\"u}r
    K{\"u}nstliche Intelligenz GmbH (DFKI)}
  \city{Saarbr{\"u}cken}
  \state{Saarland}
  \country{Germany}
}
\email{siddhant.jain@dfki.de}

\author{Anna Lea Reinwarth}
\orcid{0009-0001-7842-5928}
\affiliation{%
  \institution{Deutsches Forschungszentrum f{\"u}r K{\"u}nstliche Intelligenz GmbH (DFKI)}
  \city{Saarbr{\"u}cken}
  \state{Saarland}
  \country{Germany}
}
\email{anna\_lea.reinwarth@dfki.de}

\author{Dimitra Tsovaltzi}
\orcid{0000-0002-1670-5799}
\affiliation{%
  \institution{Deutsches Forschungszentrum f{\"u}r K{\"u}nstliche Intelligenz GmbH (DFKI)}
  \city{Saarbr{\"u}cken}
  \state{Saarland}
  \country{Germany}
}
\email{dimitra.tsovaltzi@dfki.de}

\author{Rafael Math}
\orcid{0009-0002-8633-672X}
\affiliation{%
  \institution{Deutsches Forschungszentrum f{\"u}r K{\"u}nstliche Intelligenz GmbH (DFKI)}
  \city{Saarbr{\"u}cken}
  \state{Saarland}
  \country{Germany}
}
\email{rafael.math@dfki.de}

\author{Julia Renner}
\orcid{0000-0002-4503-3178}
\affiliation{%
  \institution{National Kaohsiung University of Science and Technology}
  \city{Kaohsiung}
  \country{Taiwan}
}
\email{juliaren@nkust.edu.tw}

\renewcommand{\shortauthors}{Jain et al.}


\begin{teaserfigure}
  \centering
  \begin{tabular}{cc}
    \includegraphics[height=5cm]{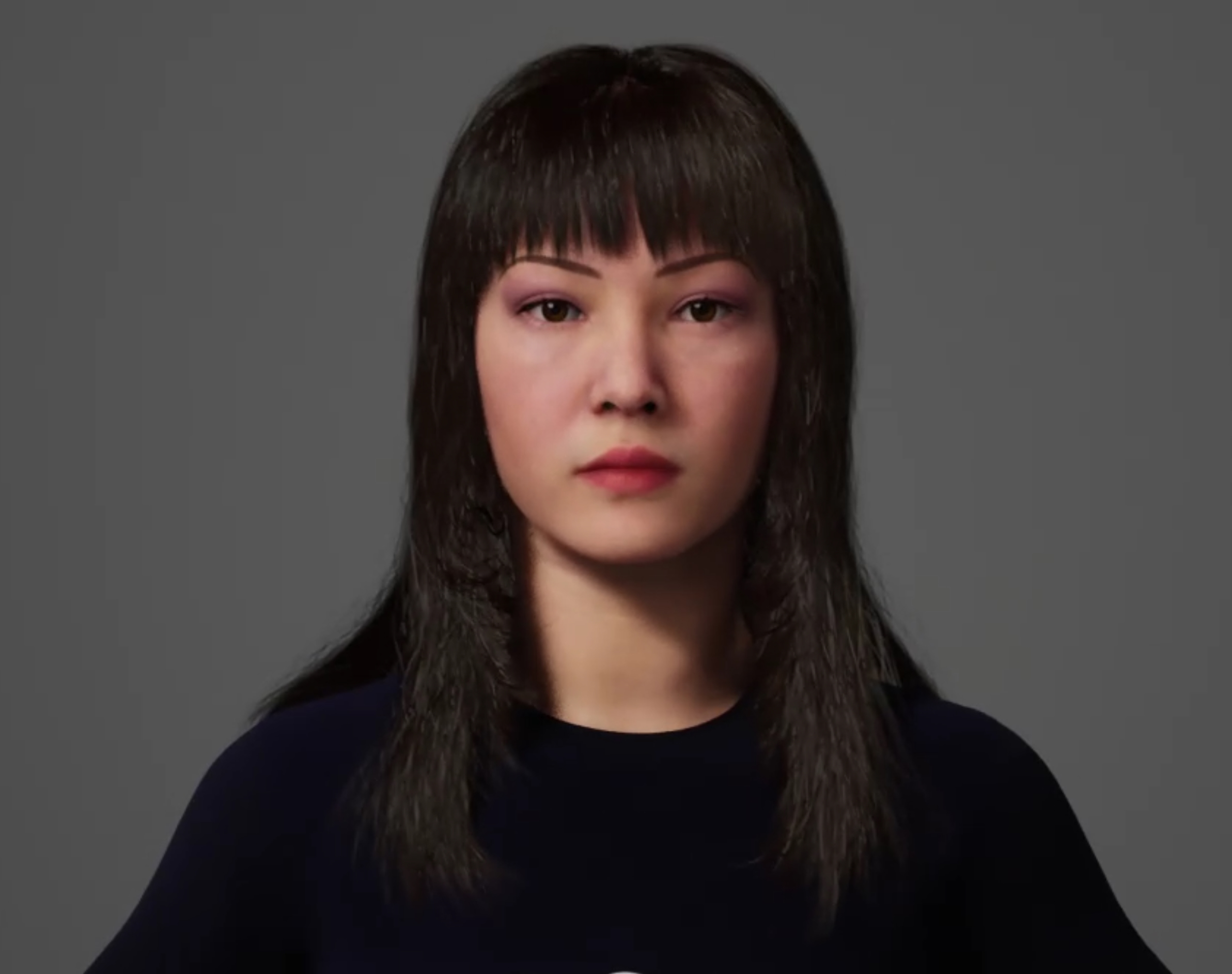} &
    \includegraphics[height=5cm]{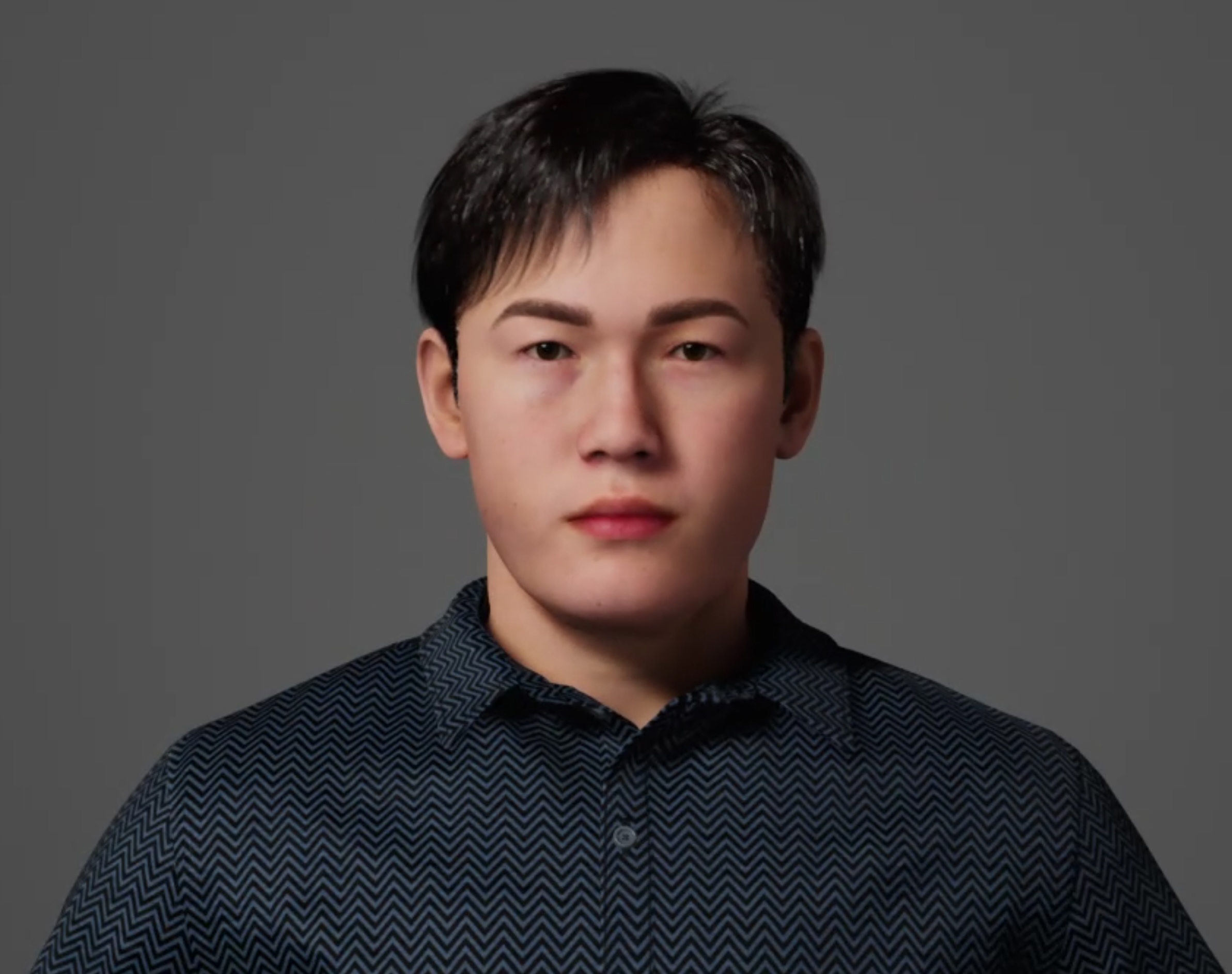} \\
    \small (a) Chen (female persona) &
    \small (b) Cheng (male persona) \\
  \end{tabular}
  \caption{The two MetaHuman personas used in the study: Chen
           (female, left) and Cheng (male, right), rendered in
           Unreal Engine 5 with real-time ARKit face animation
           and MediaPipe upper-body tracking.}
  \Description{Two photorealistic virtual human avatars rendered
               in a neutral indoor scene. The left avatar is a
               young East Asian woman named Chen. The right avatar
               is a young East Asian man named Cheng. Both have
               natural skin tones and are dressed casually.}
  \label{fig:avatars}
\end{teaserfigure}


\begin{abstract}
Successful intercultural communication requires more than
grammatical competence. It demands sensitivity to culturally
embedded social norms whose violation triggers subtle but
meaningful nonverbal responses. For German learners of Mandarin
Chinese, acquiring this sensitivity is critical yet poorly
supported by existing language-learning agents. We present a
Wizard-of-Oz (WoZ) study design and supporting real-time system
for collecting multimodal behavioral data from native Chinese
speakers reacting to social norm violations by German learners.
The system features a photorealistic MetaHuman avatar driven by
Live Link face capture and MediaPipe upper-body tracking, a wizard
console for real-time behavior selection, and synchronized
multimodal logging across agent and learner streams. A layered
annotation framework, based on psychological theory and covering
non-observable socioemotional reactions, norm interpretation,
verbal, and observable behavior thereof, and future supervision
targets enables the corpus to support training of future automated
cultural interpretation and behavior generation models. Four
ecologically valid interaction scenarios, developed with cultural
and pedagogical experts, provide the methodological and technical
foundation for a culturally adapted conversational agent for
Chinese language learning.
\end{abstract}

\ccsdesc[500]{Human-centered computing~Empirical studies in HCI}
\ccsdesc[300]{Human-centered computing~Interaction design}
\ccsdesc[300]{Applied computing~Interactive learning environments}
\ccsdesc[300]{Computing methodologies~Multimodal input}


\keywords{Embodied Conversational Agents, Wizard of Oz, Social
Norms, Socially Interactive Agents, Chinese Language Learning,
Multimodal Interaction, Intercultural Communication, Nonverbal
Behavior}

\maketitle
\pagestyle{plain}
\thispagestyle{plain}

\input{sections/01_intro}
\input{sections/02_foundations}
\input{sections/03_design}

\input{sections/04_architecture}
\input{sections/05_annotation}
\input{sections/06_discussion_ethics}


\bibliographystyle{ACM-Reference-Format}
\bibliography{references}

\end{document}

%% file: sections/01_intro.tex
\section{Introduction}

Intercultural differences and social norms affect cross-cultural communication \cite{hofstede_cultures_2001,hall_beyond_1976}. Differences in social norms can shape the nonverbal behaviors that are perceived as appropriate and polite in a situation \cite{oetzel_face_2001}. Breaking these norms can lead to miscommunications and negatively affect the quality of the relationship \cite{oetzel_face_2001}. Chinese and German communication cultures differ in several important respects, and people raised in a German context may have difficulty adapting their behavior to Chinese social norms \cite{yang_nonverbal_2013}.

\subsection{Motivation}

Culturally adapted hybrid situation aware agents can provide a low-risk training environment for German learners of Mandarin Chinese \cite{lugrin_introduction_2021,Chehayeb2025}. By carefully modeling the appropriate nonverbal behavior of a Chinese teaching situation aware agent, especially when social norms are broken, learners can become aware of their own miscommunication tendencies,  improve their cultural sensitivity and develop an understanding of how Chinese social norms differ from their own \cite{multilingual_dyadic}. However, building such an agent requires grounded behavioral data from authentic intercultural interactions, which is currently lacking.

\subsection{Research Objectives}
This research has two main objectives. First, to obtain culturally accurate verbal and nonverbal behavioral data, and facial expressions produced by native Chinese speakers socialised in a Chinese-speaking environment when social norms are broken during intercultural interactions. Second, to use these behavioural data to design and implement a culturally adapted hybrid situation aware agent as a Chinese language teacher for German-speaking learners, supported by a Wizard-of-Oz methodology \cite{woz_tools,Chehayeb2025,llm_wizards} and a four-layer multimodal annotation framework for socioemotional and behavioral data.

\subsection{Contributions}
The primary contributions of this paper are:
\begin{itemize}[leftmargin=*,itemsep=2pt,topsep=2pt]
    \item A multimodal data collection setup for capturing verbal and nonverbal responses to social norm violations in Chinese-German intercultural dialogue, built around a photorealistic situation aware agent with Live Link face capture, MediaPipe upper-body tracking, a wizard console for real-time behavior selection, and synchronized multimodal logging;
    \item A four-layer multimodal annotation framework covering non-observable socioemotional reactions, norm interpretation, verbal and observable behavior, and future supervision targets \cite{elan} for automated cultural interpretation and behavior generation.
\end{itemize}

%% file: sections/02_foundations.tex
\section{Theoretical Background}

\subsection{Hybrid Situation Aware Agents}
Hybrid situation aware agents are virtually or physically embodied systems that autonomously engage in empathetic communication with humans and other agents. They utilize a broad spectrum of multimodal behaviors, combining machine-learning-based perception with model-based reasoning about the situation to adapt their responses in real time \cite{Chehayeb2025,InCore}. The term reflects two key properties: The agents are hybrid because they combine data-driven machine learning techniques with theory-based computational models of affect and behavior (e.g., psychodynamic models), instead of relying exclusively on either approach. Furthermore, they are situation aware as they continuously monitor the state of an ongoing interaction (e.g., escalation or resolution of conflicts) and adapt their responses rather than relying only on user input \cite{Chehayeb2025,InCore}. Users frequently respond to agents in a similar way to how they respond to humans \cite{reeves_media_1996, schneeberger_fast_2023}, interpreting internal experiences through facial expressions, body movements, and vocal characteristics \cite{hortensius_perception}. Therefore, the agent design must be carefully adapted to the application context and the target user group \cite{reinwarth_look_2023}, including intercultural applications \cite{multilingual_dyadic}. Hybrid situation aware agents have been previously used in education applications \cite{Chehayeb2025}, but their success especially in a cross-cultural setting depend on the nuances of non-verbal behaviour that signals failures in communication and cannot be explicitly taught.

\subsection{Leveraging Mirroring and Marking for Nonverbal Situated Feedback in Cross-Cultural Communication}

Situation-aware agents in human-computer interaction can activate the same relational patterns as human-human interactions \cite{kircher_mentalising_2009} and can solicit more honest, open responses and an increased willingness to express negatively experienced affect \cite{lucas_computer_2014}. In cross-cultural interactions, such expression of negatively experienced affect can be intensified (see Section~\ref{sec:intercultural}), as communication may lead to perceived conflict depending on interlocutors' predispositions (see Section~\ref{sec:opd}).
 
Situation-aware agents can non-intrusively adjust their nonverbal behavior to co-regulate participants' emotional reactions in emotionally challenging interactions \cite{Chehayeb2025}, thereby eliciting expressions of negative affect. Such elicitation may involve situatively adjusted reactions perceived as negative feedback to ecologically increase perceived conflict. This way, situation-aware agents can help participants externalize negative emotions and culturally dependent interpretations, paving the way to reflection \cite{Puhl2015a}.

However, such behavior is, to our knowledge, not yet implemented in intercultural settings. One approach is through mirroring and marking in a culturally specific manner. Mirroring has been documented in caregivers who play back the infant's current facial expression \cite{taipale_self_2016} to make them aware of their emotional state. To reduce the risk of emotional misattribution, the caregiver can mark the emotion through their own expression - not mirroring fully, but altering nonverbal behavior to exaggerate or play down the emotional expressions of the infant, framing it as a reference \cite{fonagy_affect_2018}. The goal is to create an external reflection of the infant's affect-expressive display that makes it available for the infant \cite{fonagy_affect_2018}. In our case, such behavior would be used to increase awareness of one's own emotional reactions as mirrored and marked in the agent's reactions. The degree of alteration is expected to be culturally dependent, much as the intensity of emotional expressions varies across cultures. Collecting data to develop such fine-tuned yet explainable agents is the central motivation of this work.

\subsection{Intercultural Differences}
\label{sec:intercultural}
Chinese and German communication are embedded in distinct cultural traditions \cite{hofstede_cultures_2001,hall_beyond_1976} that shape social norms, nonverbal behavior, and expression rules across all aspects of interaction \cite{yang_nonverbal_2013}. Cross-cultural research distinguishes grammatical errors from social norm violations, the latter being more consequential as interlocutors tolerate grammatical mistakes more readily \cite{oetzel_face_2001}. Following Hofstede's cultural dimensions and Hall's high- and low-context framework, China is associated with higher power distance, stronger collectivist orientations, and higher-context communication, whereas Germany favors lower power distance, individualism, and explicit, direct argumentation \cite{hofstede_cultures_2001,hall_beyond_1976}. These differences are reflected in conflict management \cite{oetzel_face_2001} and in everyday behavioral scripts such as gift-giving practices \cite{meng_social_2008}. A complementary perspective is offered by the IAHA Cultural Responsiveness Framework \cite{iaha}, which conceptualizes cultural responsiveness as a way to address cross-cultural communication, moving from mere cultural knowledge toward actively responsive, respectful engagement, structured around dimensions such as self-awareness, relational awareness, and power-and-context awareness.

In hybrid situation aware agent research, cultural adaptation is central and should be based on an awareness of cultural differences, models of addressing cross-cultural communication. They target that users infer an agent's cultural background from its behavior \cite{lugrin_introduction_2021}, favor agents tailored to their own culture \cite{zhou_adapting_2017}, and perceive agent feedback differently depending on their cultural background \cite{wang_wow_2020}. Computational analysis of Asian-European dyadic interaction confirms that nonverbal features vary systematically across cultures and influence engagement \cite{multilingual_dyadic}. These intercultural differences in emotional expression and conflict behavior are further grounded in psychodynamic constructs, particularly OPD lead affect and internal conflict patterns which provide the explainability layer connecting nonverbal behavior to underlying motivational states (Section~\ref{sec:opd}).

\subsection{Psychodynamic Foundation}
\label{sec:opd}
Operationalized Psychodynamic Diagnostics (OPD) is a structured framework, originally developed for psychotherapy, that describes a person's recurring relational patterns rather than isolated symptoms \cite{OPD}. It is aimed to capture affect reactions in everyday situations; psychoanalysis is in general used to explain not only psychological pathologies, but also everyday life situations. OPD assumes that behavior is driven by a finite set of stable, generalizable internal conflicts (e.g., need to control vs. comply), and one lead affect (e.g., fear) dominates and influences how a person acts to manage that conflict \cite{OPD}. Because these conflicts recur across interactions, OPD offers a compact, theory-grounded framework for modeling the affective and motivational aspects of hybrid situation aware agents, especially when modeling conflict escalation and resolution in intercultural interactions.

\subsection{Wizard-of-Oz for Multimodal Data Collection}
The Wizard-of-Oz (WoZ) method is an established technique in which a human operator simulates a not-yet-automated system while participants interact naturally \cite{woz_tools}. WoZ is particularly suited to early-stage agent research where naturalistic behavioral data must be collected before automation is feasible \cite{llm_wizards}, and has been used in language learning contexts to elicit authentic learner responses \cite{woz_language_learning}. For multimodal research it offers full control over the agent's verbal and nonverbal output, which may align with different communication and socioemotional models, for example through training the wizards or through predefining possible decision paths, while capturing unconstrained learner behavior. The wizard's decision trace and accompanied spontaneous nonverbal behavior provides ground-truth labels for training future cultural interpretation and behavior selection models \cite{llm_wizards,woz_tools,multilingual_dyadic}.
 
The agent acts within a prestructured communication setting where interaction scenarios are chosen for the kind of internal conflicts they may provoke in the specific intercultural situation and based on individual predispositions, following the OPD framework. Culturally dependent social norms are identified and their cross-cultural interpretation is collected using qualitative methods applied during post-session annotation. Recent work has explored using large language models as wizards to scale WoZ experiments \cite{llm_wizards} and cooperative multi-wizard platforms to distribute the cognitive load of real-time behavior control \cite{wizundry}. In our study, the wizards have expertise in the cultural norms being studied, ensuring that agent responses are both linguistically authentic and culturally grounded.

\begin{figure*}[t]
  \centering
  \includegraphics[width=\textwidth, height=12cm, keepaspectratio]{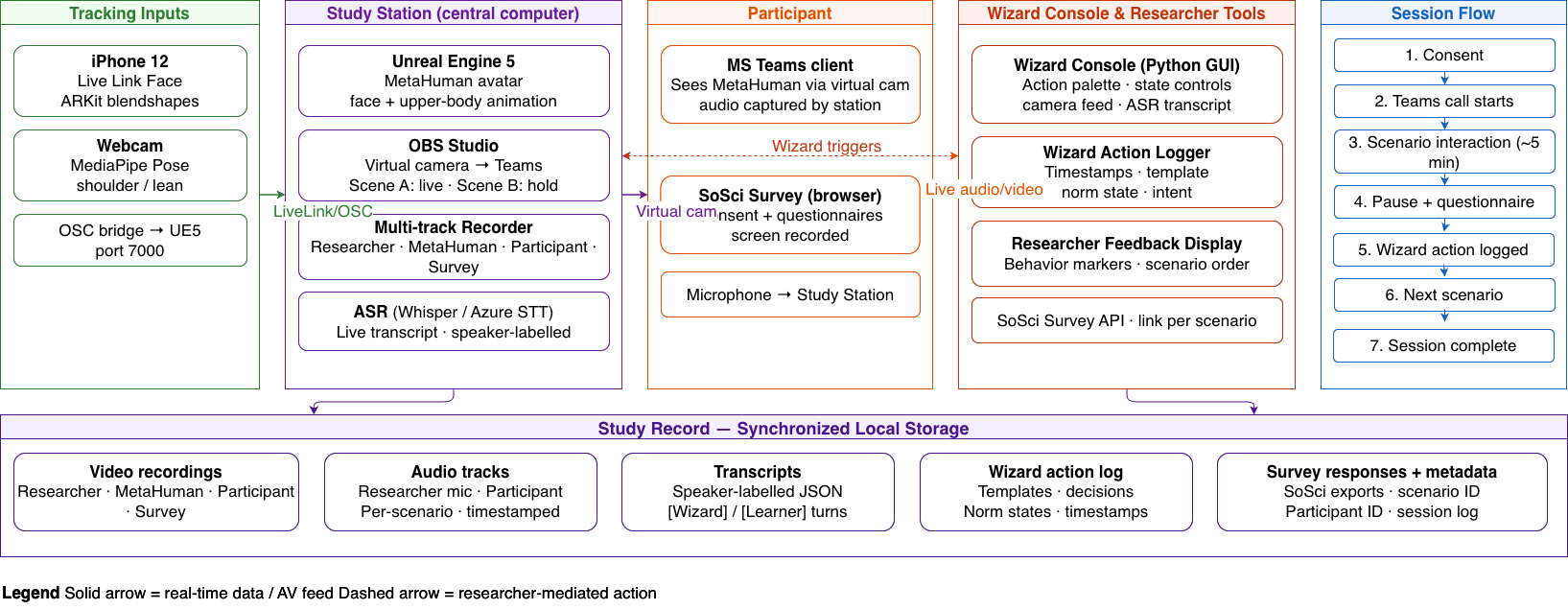}
  \caption{End-to-end WoZ study setup. Solid arrows indicate real-time data and AV feeds. Dashed arrows indicate researcher-mediated actions. The wizard console triggers avatar behavior templates in real time while all streams are synchronized and logged for post-session annotation.}
  \Description{A system architecture diagram divided into five zones: tracking inputs showing iPhone and webcam; study station showing Unreal Engine, OBS, multi-track recorder and ASR; participant zone showing MS Teams and SoSci Survey; researcher tools showing Python controller, wizard console, and feedback display; and a study record zone showing all stored data streams. A session flow column on the right shows eight steps from consent through session completion.}
  \label{fig:setup}
\end{figure*}

%% file: sections/03_design.tex
\section{Study Design}

\subsection{Research Questions}
The study is guided by three research questions:

\begin{description}[leftmargin=0pt,itemsep=2pt,topsep=2pt]
\item[\textbf{RQ1}] To what extent can a WoZ methodology with a photorealistic hybrid situation aware agent elicit ecologically valid intercultural interaction data for future automated behavior modeling?
\item[\textbf{RQ2}] Which verbal and nonverbal behaviors do Chinese speakers produce in response to social norm violations by German learners during intercultural dialogue?
\item[\textbf{RQ3}] Which multimodal response patterns of the Chinese interlocutor vary across different types of social norm violation?
\end{description}

\subsection{Interaction Scenarios}
Four interaction scenarios were developed in close cooperation with psychology, learning science, and Chinese language teaching experts. Each scenario depicts a realistic social situation in which intercultural norm conflicts between German and Chinese conventions are likely to arise. The scenarios are designed to provoke the kinds of internal conflicts described by the OPD framework \cite{OPD}, such as the tension between autonomy and dependency, or submission and control that are expected to surface in Chinese-German intercultural encounters. Each scenario is annotated for socioemotional response, norm violation type, and markedness level \cite{fonagy_affect_2018, oetzel_face_2001}, providing the theoretical grounding needed to connect observable nonverbal behavior to underlying motivational and affective states. The scenarios are presented in a fixed order, progressing from social ambiguity to clear norm violation, and concluding with an affirmative baseline condition designed to collect negative training examples (i.e., agent behavior when norms are upheld).

\input{tables/scenarios}

The scenarios are not randomized: the fixed sequence allows us to observe conflict escalation and recovery across a session. Sub-scenario variations (e.g., whether a learner accepts or declines food and drink) are not controlled, preserving ecological validity and yielding naturalistic behavioral variance. The final scenario (S4) uses an affirmative agent response regardless of learner behavior, providing contrastive data for model training.

\subsection{Participants}
The study involves two participant roles. The \textit{learner} is a native German speaker with intermediate or advanced Mandarin proficiency (B1--C1 CEFR). Two \textit{wizards} participate in the study, both trained in the WoZ protocol. An important criterion for selecting wizards is that they must be socialised in a Chinese-speaking environment to ensure that the authenticity of nonverbal behavior spontaneously carries over into the interaction. A pre-study questionnaire assesses the learner's familiarity with Chinese cultural norms through open questions, to avoid priming effects. The exact sample sizes will be determined by power analysis; we target a minimum of 20 learner sessions to ensure sufficient behavioral variation between scenarios.

\subsection{Experimental Protocol}
Each session follows a fixed measurement sequence:

\smallskip
\noindent\textit{Baseline check} $\rightarrow$ S1 $\rightarrow$ Feedback $\rightarrow$ S2 $\rightarrow$ Feedback $\rightarrow$ S3 $\rightarrow$ \textit{Peak check} $\rightarrow$ Feedback $\rightarrow$ S4 $\rightarrow$ Feedback
\smallskip

\noindent A brief self-report measure captures the learner's affective and relational state at the start and after the highest-conflict scenario. Between scenarios, the OBS scene switches to a hold screen and a short debrief questionnaire is sent to the learner via the Teams chat, inviting reflection on the interaction without revealing the cultural norm focus of the scenario.
 
The session is managed through the wizard console, which tracks the current state machine position and prompts the wizard with the next step at each transition. This reduces cognitive load during the interaction and ensures the fixed measurement sequence is maintained across all sessions. All console actions are timestamped automatically, producing the wizard action log used in Layer C annotation.

\subsection{Data Collection}
Each session produces four synchronized data streams:

\begin{itemize}[leftmargin=*,itemsep=2pt,topsep=2pt]
    \item \textbf{Agent stream}: Triggered behavior templates with timestamps and wizard action labels. Facial blendshape values (52 ARKit parameters at 60\,fps) are logged for display documentation and are not used as training data.
    \item \textbf{Learner stream}: RGB video from a secondary camera, close-up audio, and ASR transcript (automated, cleaned by linguists post-session).
    \item \textbf{Wizard log}: Decision state, selected behavior template, and timestamp for each wizard action.
    \item \textbf{Session metadata}: Scenario identifier, participant ID, WoZ decision trace, and questionnaire responses.
\end{itemize}

\subsection{Planned Analysis}
Post-session annotation will follow a four-layer schema (Section~\ref{sec:annotation}). Descriptive analysis will characterize the distribution of agent and learner behaviors across scenarios and types of norm violation, including markedness level and OPD conflict category. Cross-wizard consistency of behavior template selection will be assessed to distinguish individually idiosyncratic responses from broadly representative cultural patterns, addressing the potential subjectivity inherent in the 
wizard role. Future work will use the annotated corpus to train automated components - a cultural interpretation model and a behavior selection policy that progressively replace the wizard in the runtime pipeline.

%% file: tables/scenarios.tex
\begin{table}[h]
\caption{Interaction scenarios targeting Chinese-German social norm differences. S4 collects contrastive data by affirming all learner behavior.}
\label{tab:scenarios}
\centering
\small
\begin{tabular}{p{0.5cm}p{1.8cm}p{2.8cm}p{2.0cm}}
\toprule
\textbf{ID} & \textbf{Scenario} & \textbf{Norm Focus} & \textbf{Strategy} \\
\midrule
S1 & Dinner invitation & Food/drink refusal without accepted reason & Nonverbal discomfort \\
\addlinespace
S2 & Restaurant invitation & Accepting without resistance when host implicitly invites & Mild social tension \\
\addlinespace
S3 & Declining an invitation & Direct refusal without face-saving & Responsive \\
\addlinespace
S4 & Presentation planning & Indirect disagreement among peers & Affirm all behavior \\
\bottomrule
\end{tabular}
\end{table}

%% file: sections/04_architecture.tex
\section{System Architecture}
\label{sec:system}

The study is conducted fully online: the learner participates remotely via Microsoft Teams, while the wizard and study computer operate from a central location. The WoZ system connects four physical components - a tracking station, a study computer, a participant-facing interface, and a researcher console into a synchronized pipeline for real-time interaction and multimodal data capture (Figure~\ref{fig:setup}). The MetaHuman avatar serves exclusively as the \textit{presentation layer}, rendering the culturally appropriate agent behavior selected by the wizard. No avatar rendering data are used for machine learning; the corpus for future modeling consists of annotated learner behavior signals and wizard decision traces only.

\subsection{Tracking and Avatar Realization}
Face animation is driven in real time by an iPhone~12 via Apple ARKit and Unreal Live Link, capturing 52 facial blendshape parameters at 60\,fps. Upper-body motion - shoulder sway and forward lean - is tracked from a laptop webcam using MediaPipe Pose \cite{mediapipe} and transmitted to Unreal Engine~5 via OSC. The MetaHuman avatar \cite{highfidelity_avatar} is rendered on the study computer and streamed to the learner through OBS Studio as a virtual camera feed over Microsoft Teams, preserving ecological validity for a remote tutoring context. Audio is delivered by a pre-recorded or live text-to-speech system selected by the wizard.

\subsection{Offline Authoring}
Before each session, domain experts prepare: (1) \textit{scenario scripts} encoding cultural context and expected norm-violation moments; (2) \textit{persona configurations} specifying agent identity and cultural background; (3) \textit{a behavior library} of pre-authored animation templates for facial expressions, gaze, gestures, and posture, grounded in documented Chinese nonverbal behavior \cite{yang_nonverbal_2013,chui_language_2022,reinwarth_look_2023}; and (4) \textit{a cultural knowledge base} encoding relevant social norms and OCC-based appraisal labels \cite{occ_model}.

\subsection{WoZ Console and Logging}
\begin{figure}[h]
  \centering
  \includegraphics[width=\columnwidth]{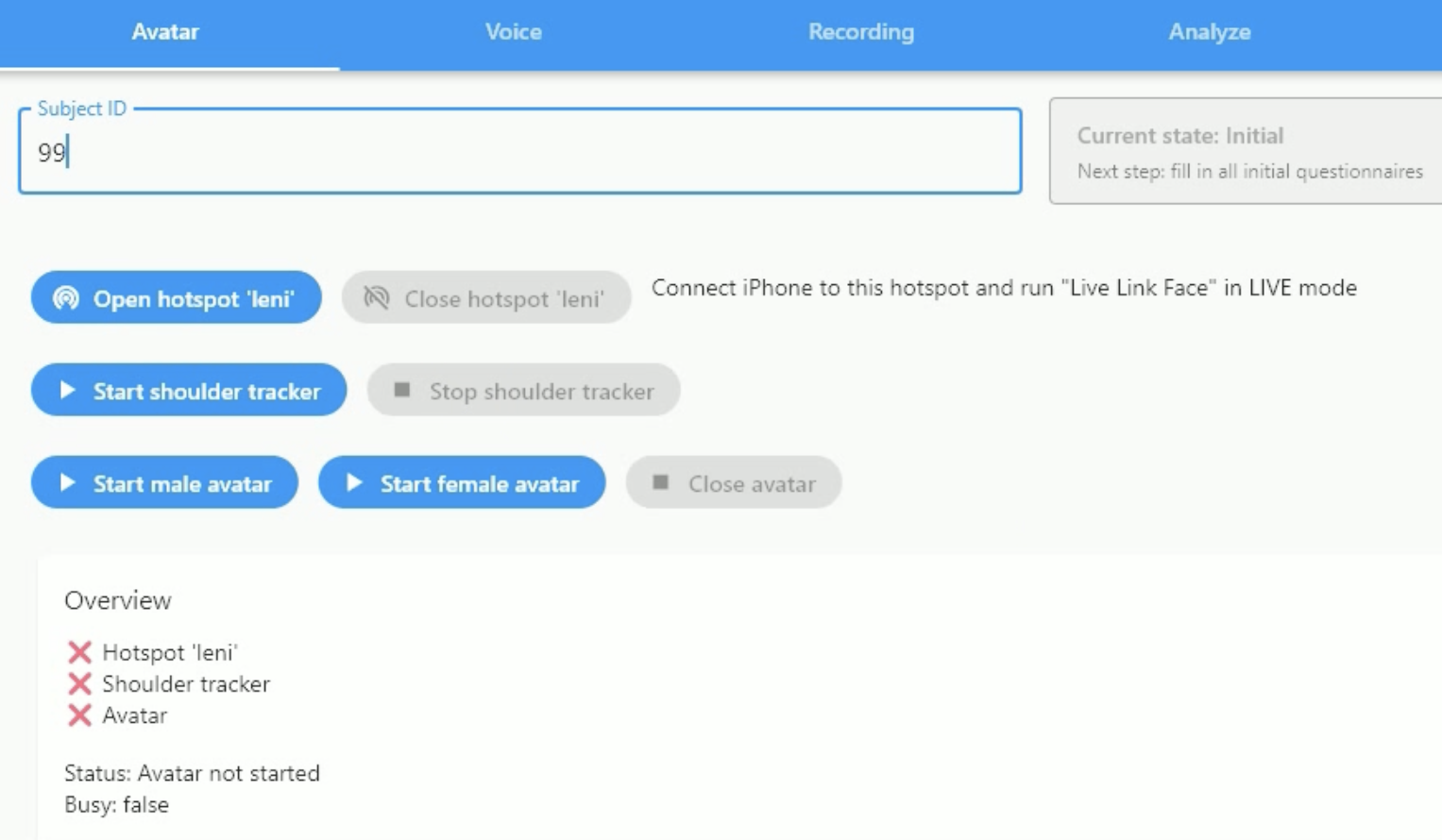}
  \caption{Wizard console Avatar tab showing subject ID entry, iPhone
           hotspot control for Live Link face capture, shoulder tracker
           launch, and avatar persona selection. Status indicators confirm
           system readiness before each session begins.}
  \Description{Screenshot of the wizard console Avatar tab. Shows a
               Subject ID field, buttons for opening a hotspot, starting
               the shoulder tracker, and launching male or female avatars.
               An overview panel shows red crosses next to hotspot,
               shoulder tracker, and avatar indicating they are not yet
               started.}
  \label{fig:console}
\end{figure}

The wizard monitors the remote learner through a live video feed and updated ASR transcript in real time (Figure~\ref{fig:console}). A categorized action palette allows the wizard to trigger behavior templates on the avatar instantly. Each selection writes a timestamped entry to the wizard action log recording the selected template and inferred norm state. Communicative intention labels (e.g., \textit{acknowledge}, \textit{express mild disapproval}, \textit{redirect}, \textit{affirm}) are assigned during post-session annotation.

All streams - learner video, ASR transcript, and wizard action log are stored with a shared session timestamp, enabling frame-accurate alignment during annotation. A multi-track recorder captures synchronized audio and video across all session participants simultaneously. Between scenarios, the OBS scene switches to a hold screen while the learner completes a brief online questionnaire via SoSci Survey, sent through the Teams chat.

\begin{figure}[t]
  \centering
  \includegraphics[width=\columnwidth]{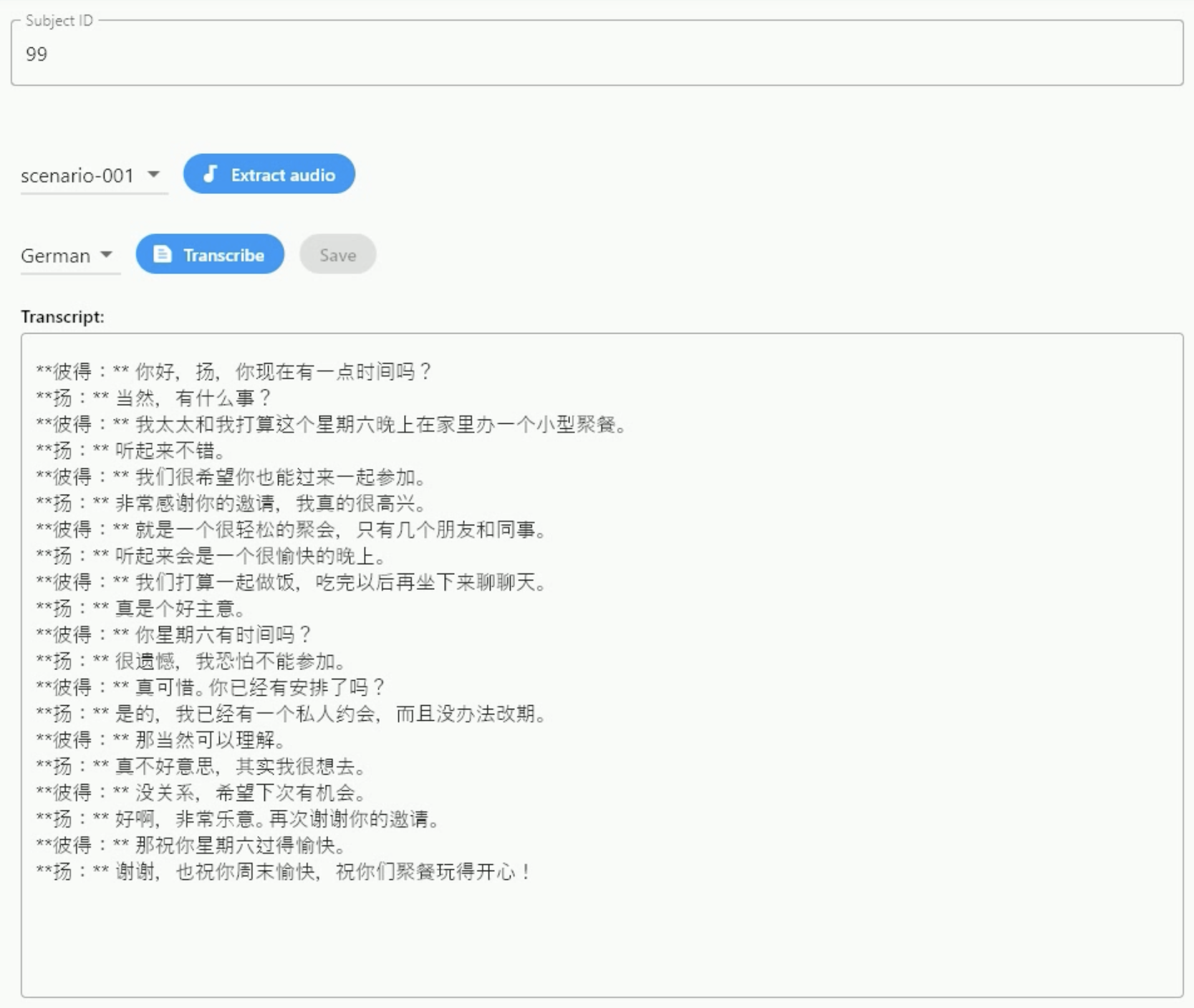}
  \caption{Wizard console Transcript tab showing the ASR-generated
           Chinese dialogue transcript with speaker labels. Transcripts are stored
           automatically during the session and used for
           post-session annotation.}
  \Description{Screenshot of the wizard console Transcript tab
               showing subject ID 99, a scenario selector set to
               scenario-001, extract audio and transcribe buttons,
               and a Chinese language transcript with speaker
               labels showing a dialogue between two speakers
               about a dinner invitation.}
  \label{fig:transcript}
\end{figure}

%% file: sections/05_annotation.tex
\section{Annotation Framework}
\label{sec:annotation}

The annotation schema is organized into four independent layers, allowing different annotators to work in parallel and enabling the corpus to support multiple modeling tasks.

\textbf{Layer A - Observable Behavior} captures time-aligned, interpretation free labels for both the agent and the learner: speech act type (accept, mitigate, refuse, explain, delay, politeness marker), facial expression category, gaze direction, upper-body gesture (e.g., hand block, accepting gesture, hesitation), and postural shift (lean forward/back, shoulder tension). Agent-side annotations include the triggered behavior template identifier and the wizard action label.
 
\textbf{Layer B - Norm Interpretation and Socioemotional State} assigns each interaction segment a social obligation label (e.g., greeting, request, gratitude, leave-taking), a violation judgment (adhered / borderline / violated), a markedness rating \cite{fonagy_affect_2018,oetzel_face_2001}, and the cultural norm category from the knowledge base. Non-observable socioemotional reactions - including perceived discomfort, face threat, and relational tension inferred from the interaction context - are annotated at this layer, bridging observable behavior (Layer A) and the psychodynamic constructs used in Layer D.
 
\textbf{Layer C - Wizard Decisions and Mirroring-Marking Labels} records, for each wizard action, the selected behavior template and the communicative intention assigned post-session. Each wizard action is further labeled for mirroring type (full mirror, marked alteration, or no mirror) and marking degree (exaggerated, attenuated, or neutral), reflecting the extent to which the agent's nonverbal response reflects and reframes the learner's affective display \cite{fonagy_affect_2018,taipale_self_2016}. The inferred conflict state follows the OPD framework \cite{OPD}. This layer directly supports the training of a future automated policy for culturally grounded nonverbal behavior generation.
 
\textbf{Layer D - Learning Targets} captures higher-level constructs derived from the full session: learner repair attempts, main affective reaction (OPD lead affect mappings on OCC emotions \cite{occ_model}), interpersonal conflict indicators, and conflict escalation trajectory. These labels serve as supervision targets for future cultural interpretation and behavior generation models.

Annotation will be performed in ELAN \cite{elan}, using a shared tier schema exported to the corpus alongside the raw media files. Inter-annotator agreement will be computed using Cohen's $\kappa$ for categorical labels and intraclass correlation for continuous ratings.

%% file: sections/06_discussion_ethics.tex
\section{Discussion and Future Work}

The proposed WoZ methodology offers several advantages for collecting culturally grounded multimodal interaction data. The fixed scenario sequence controls contextual variation while preserving behavioral naturalness, and the photorealistic avatar is designed to reduce the uncanny valley effect that can suppress authentic learner responses
\cite{highfidelity_avatar}. The layered annotation schema decouples observable behavior from interpretive labels, making the corpus reusable across tasks from norm detection to behavior generation. The integration of OPD-based conflict constructs and mirroring-marking labels into the annotation framework further ensures that the corpus supports not only
behavioral modeling but also theoretically grounded, explainable agent design.

\subsection{Training Plan}
The annotated corpus will support the training of three automated modules that progressively replace the wizard in the runtime pipeline. First, a \textit{cultural interpretation model} will classify social norm adherence, violation type, and markedness level from multimodal learner signals combining facial expression, gaze, posture, and ASR transcript features. Second, a \textit{dialogue orchestrator} will select agent responses
conditioned on the inferred norm state, OPD conflict category, conflict escalation potential \cite{InCore}, and markedness level, trained on wizard decision traces from Layer C. Third, a \textit{behavior generation model}
will produce avatar animation from selected response templates, using the mirroring and marking labels from Layer C as fine-grained supervision targets. Together, these components form the automated hybrid situation aware
agent, with the WoZ layer removed from the runtime pipeline \cite{Chehayeb2025,InCore}.
 
\subsection{Evaluation Plan}
Evaluation will proceed in two stages. In the first stage, model performance will be assessed automatically using the annotated corpus: the cultural interpretation model will be evaluated on classification accuracy for norm violation type and markedness level against Layer B and Layer D ground truth; the dialogue orchestrator will be evaluated on action selection accuracy against Layer C wizard decisions. In the second stage, a user study will assess the fully automated agent in terms of perceived cultural authenticity, ecological validity, and learner reflection outcomes, comparing agent behavior with and without the mirroring-marking component. Expert raters with Chinese intercultural communication expertise will provide qualitative validation of agent nonverbal responses against the behavior library.

\subsection{Reproducibility and Open Science}
To support reproducibility and transparency, the wizard console, MediaPipe bridge, and avatar configuration scripts will be made available as open-source tools. The annotated corpus will be released under a data sharing agreement following pseudonymization, subject to ethical approval and participant consent. The ELAN tier schema and behavior library taxonomy will be published alongside the corpus to enable reuse in related intercultural agent research.

\subsection{Limitations}
As a pre-data collection paper, empirical validation remains a future work. The wizard role introduces subjectivity into behavior selection; the use of two wizards, combined with the auditable action log and cross-wizard consistency analysis, mitigates individual bias. The scenario set targets Chinese-German dyadic interaction and may require adaptation for other intercultural pairs. The online study setup via Microsoft Teams, while ecologically valid for remote tutoring, may affect the naturalness of nonverbal behavior compared to face-to-face interaction, particularly for gaze
direction, interpersonal distance, and gesture range, which are constrained by camera framing and screen mediation \cite{multilingual_dyadic}.

\section{Ethics}

The study protocol has been submitted for institutional ethical review. Participants will provide their written informed consent for audio-video recording and data use. Data will be pseudonymized and stored on GDPR-compliant servers. The wizard role will be disclosed during the post-study debriefing, consistent with standard WoZ practice \cite{woz_tools}\footnote{Following current Open Science recommendations, the WoZ user study will be preregistered on OSF prior 
to data collection.}.